\documentclass[doublespacing]{revtex4-1}
\usepackage{natbib}
\usepackage{tikz}
\usepackage{amssymb}
\usepackage{amsmath}
\usepackage{graphics}
\usepackage{graphicx}
\usepackage{color}
\usepackage{rotating}
\newcommand{\dspaceon}{\renewcommand{\baselinestretch}{2.0}\large\normalsize}

\makeatletter
\newsavebox\myboxA
\newsavebox\myboxB
\newlength\mylenA
\newcommand*\xoverline[2][0.75]{%
    \sbox{\myboxA}{$\m@th#2$}%
    \setbox\myboxB\null% Phantom box
    \ht\myboxB=\ht\myboxA%
    \dp\myboxB=\dp\myboxA%
    \wd\myboxB=#1\wd\myboxA% Scale phantom
    \sbox\myboxB{$\m@th\overline{\copy\myboxB}$}%  Overlined phantom
    \setlength\mylenA{\the\wd\myboxA}%   calc width diff
    \addtolength\mylenA{-\the\wd\myboxB}%
    \ifdim\wd\myboxB<\wd\myboxA%
       \rlap{\hskip 0.5\mylenA\usebox\myboxB}{\usebox\myboxA}%
    \else
        \hskip -0.5\mylenA\rlap{\usebox\myboxA}{\hskip 0.5\mylenA\usebox\myboxB}%
    \fi}
\makeatother

\def \a{\`{a}}

\def \i{\`{\i}~}

\def \d{\mbox{d}}

\def\centerarc[#1](#2)(#3:#4:#5){ \draw[#1] ($(#2)+({#5*cos(#3)},{#5*sin(#3)})$) arc (#3:#4:#5); }
\begin{document}

%==================
%\begin{frontmatter}
%==================

% Use optional labels to link authors explicitly to addresses:

\title{
%Ductile breakage rate of small aggregates by hydrodynamic
%stress\\
%in anisotropic turbulence
Turbulent breakage of ductile aggregates
}

\vspace{-1.0cm}
\author{
Cristian Marchioli\footnote{Author to whom correspondence
should be addressed E-mail: marchioli@uniud.it~~Phone: +39~(0)432 558020.}
and Alfredo Soldati
}
% To be used with revtex
%\footnote{Corresponding author. E-mail: soldati@uniud.it,
%Tel.: +39 0432 558020}, M. Campolo and C. Marchioli}
% To be used with elsevier
%\corauth[cor1]{Corresponding author.
%Also affiliated with Department of Fluid
%Mechanics, CISM, 33100 Udine, Italy.
%E-mail: marchioli@uniud.it, Tel.: +39 0432 558020.
%}

\address{
Dipartimento di Ingegneria Elettrica, Gestionale e Meccanica,
Universit\a~di Udine, Udine (Italy)\\
Dipartimento di Fluidodinamica, CISM, Udine (Italy)
}

%-------------------------------------------------------
%
%\makeatletter
%\global\@specialpagefalse
%\def\@oddhead{ \ifnum\c@page=1
%\let\@evenhead\@oddhead
%\llap{ \hfill Submitted to: {\em Int. J. Multiphase Flow} }
%\fi }
%\let\@evenfoot\@oddfoot
%\makeatother
%
%-------------------------------------------------------

\begin{abstract}
In this paper we study breakage rate statistics of small colloidal aggregates
in non-homogeneous anisotropic turbulence.
We use pseudo-spectral direct numerical simulation of
turbulent channel flow and Lagrangian tracking to follow the motion
of the aggregates, modelled as sub-Kolmogorov massless particles.
We focus specifically on the effects produced by ductile rupture:
This rupture is initially activated when fluctuating hydrodynamic stresses
exceed a critical value, $\sigma>\sigma_{cr}$,
and is brought to completion when the energy absorbed by the aggregate meets
the critical breakage value.
We show that ductile rupture breakage rates are significantly
reduced with respect to the case of instantaneous
brittle rupture (i.e. breakage occurs as soon as $\sigma>\sigma_{cr}$). These discrepancies
are due to the different energy values at play as well as
to the statistical features of energy distribution in the anisotropic
turbulence case examined.
\end{abstract}

\maketitle

\dspaceon

%\begin{keyword}
%Turbulent Boundary Layer \sep
%Coherent Structures \sep
%Particle Dispersion \sep
%Direct Numerical Simulation \sep
%Lagrangian Particle Tracking \sep
%Deposition \sep
%Preferential Distribution \sep
%Re-entrainment.
%\end{keyword}

%================
%\end{frontmatter}
%================

%%%%%%%%%%%%%%%%%%%%%
\section{Introduction}
%%%%%%%%%%%%%%%%%%%%%

Breakage rates of micro and nano aggregates in turbulent flow are of high relevance
to a wide variety of applications. These range from industrial processes,
such as operations involving colloids \cite{soos1}, compounding of
plastic and elastomeric materials \cite{yin07}, and dispersion of ceramics \cite{malghan},
to environmental processes, e.g. sedimentation of marine snow \cite{burd}
or formation of marine aggregates \cite{ruiz}. In almost all cases
of practical interest, breakage is caused by two mechanisms.
The first is usually referred to as impact breakage and is caused by energetic
collisions of the aggregates with other aggregates or walls. This mechanism, however,
is not relevant to the present study because it applies to aggregates
that are large with respect to the characteristic length scale of the fluid
shear (the Kolmogorov length scale if the flow is turbulent)
and have a significant density difference compared to the fluid.
The second mechanism is usually referred to as shear breakage and applies to small
aggregates with density close to that of the fluid.
In this case, breakage is caused by
aggregate deformation induced by the hydrodynamic stresses.
Such deformation generates internal stresses that can break the aggregate following
two processes:
If the response time of the aggregate to deformation is very
small then breakage is instantaneous, and the aggregate can be referred to as brittle;
if breakage depends on the stress history and a significant amount of energy
is required to overcome deformation, then the aggregate can be referred to as ductile.

In either brittle or ductile rupture, the phenomenology of turbulent breakage is
still not fully understood because the complexity of the flow field
adds to the intricacy of the aggregate morphology in determining how the hydrodynamic
forces redistribute over the structure of the aggregate and how stresses accumulate
in critical locations
where the cohesive force that keeps the primary particles of the aggregate together can be overcome.
To provide a basic understanding of turbulent breakage, many investigations (see for
instance the recent works by \cite{thesJFM,lanotte-vanni,lanotte,thes2008}
%Babler et al. (2008, 2012, 2015), De Bona et al. (2014)
and the references therein)
have focused on the influence that the hydrodynamic stresses have on the rate at which breakage occurs,
neglecting the details of the aggregate inner structure.
This choice \cite{footnote1}
was motivated by the observation that the occurrence, in the surroundings of a small aggregate,
of instantaneous stresses capable of inducing breakage is controlled by the statistical (spatial and temporal)
distribution of these stresses, which is in turn dictated by the turbulent flow field \cite{lanotte-vanni}.
However, in the size range of interest for the present study (aggregates
smaller than the Kolmogorov length),
analyses were carried out considering brittle aggregates and instantaneous
breakage \cite{thesJFM,lanotte}.
%(Babler et al., 2012, 2015).
This assumption would be fully justified in highly viscous flows, such as dispersions
in liquid polymers, where the stresses required to break the aggregate
are very low \cite{vanni}.
But in low-viscosity systems the effective hydrodynamic stress required for breakage
changes significantly depending on the nature of the flow
and rupture can frequently be determined by the stress history \cite{blaser}, namely by what we refer to as
ductile rupture effects in this paper.
As a result, current knowledge of the breakage process for the case of sub-Kolmogorov ductile aggregates
has remained hitherto elusive in spite of its practical importance in areas such as industrial
materials processing \cite{bumm} and rheology of dense colloids \cite{zaccone}.
Design of such processes usually relies on semi-empirical correlations based on
idealized flow configurations \cite{gv2011,vg2011,soos2010,soos2008,kust96}.
%(Gastaldi \& Vanni, 2011; Vanni \& Gastaldi, 2011; Soos et al., 2010;
%Soos et al., 2008; Babler et al., 2008; Kusters et al., 1996).
When turbulence is present, it is generally modelled
with a single value of the kinetic energy dissipation, a rather crude assumption
already for homogeneous and isotropic flows. However, real facilities involve mixing
in containers, agitators, flows in pipes and channels where turbulence is strongly
anisotropic and geometry-dependent \cite{soos2010,soos2008,kust96,kust}.
Compared to the few studies of the influence of turbulence in homogeneous and isotropic flow
conditions (see \cite{lanotte-vanni} and \cite{lanotte} among others), the first attempt
to assess the effects of flow inhomogeneity and anisotropy was recently put forward in
the collaborative study of \cite{thesJFM}. In this study, direct numerical simulations
were performed to study the breakage of small inertialess aggregates in different archetypal
bounded flows
as turbulent channel flow (data and analysis provided by this group) and developing boundary layer,
comparing results with those of homogeneous isotropic turbulence.
To emphasize the dependence of breakage on the specific properties of the flow, only brittle instantaneous
rupture was considered in the simulations.
The main findings, relevant to the present
discussion, can be summarized as follows: (I) Regardless of the flow configuration, the breakage
rate decreases when the critical stress required to break the aggregate increases; (II) for small
values of the critical stress (``weak'' aggregates) the breakage rate develops a universal power-law
scaling that appears to be independent of the flow configuration; (III) for high values of the critical stress
(``strong'' aggregates) large differences in the breakage rate arise among the different flows and no
clear scaling is observed anymore, highlighting the importance of instantaneous and local flow properties.

Different from \cite{thesJFM}, in the present paper we examine a more realistic breakage process
that results from ductile (non-instantaneous) rupture, focusing on the case of turbulent channel flow.
We are interested in
assessing the influence of ductile rupture on the breakage rate,
with specific reference to the scaling behavior previously observed
for weak aggregates \cite{thesJFM,lanotte}.
The statistical characterization proposed in this work also provides useful
indications about the importance of ductile rupture in the context of Large-Eddy
Simulations (LES) of industrial particulate systems. More specifically, the
results discussed here suggest that the specific breakage
mechanism has a crucial impact on the modelling framework within which LES can be
used. As mentioned, shear breakage in anisotropic turbulence is controlled by
the fluid stresses around the aggregate but these stresses are inevitably
filtered out by LES, thus causing under-prediction of the breakage rates.
This under-prediction is expected to be more evident in the case of ductile
rupture because the contribution of subgrid fluctuations to the
stress history is missing.

\begin{figure}
\begin{center}
\includegraphics[width=10.0cm,angle=0.]{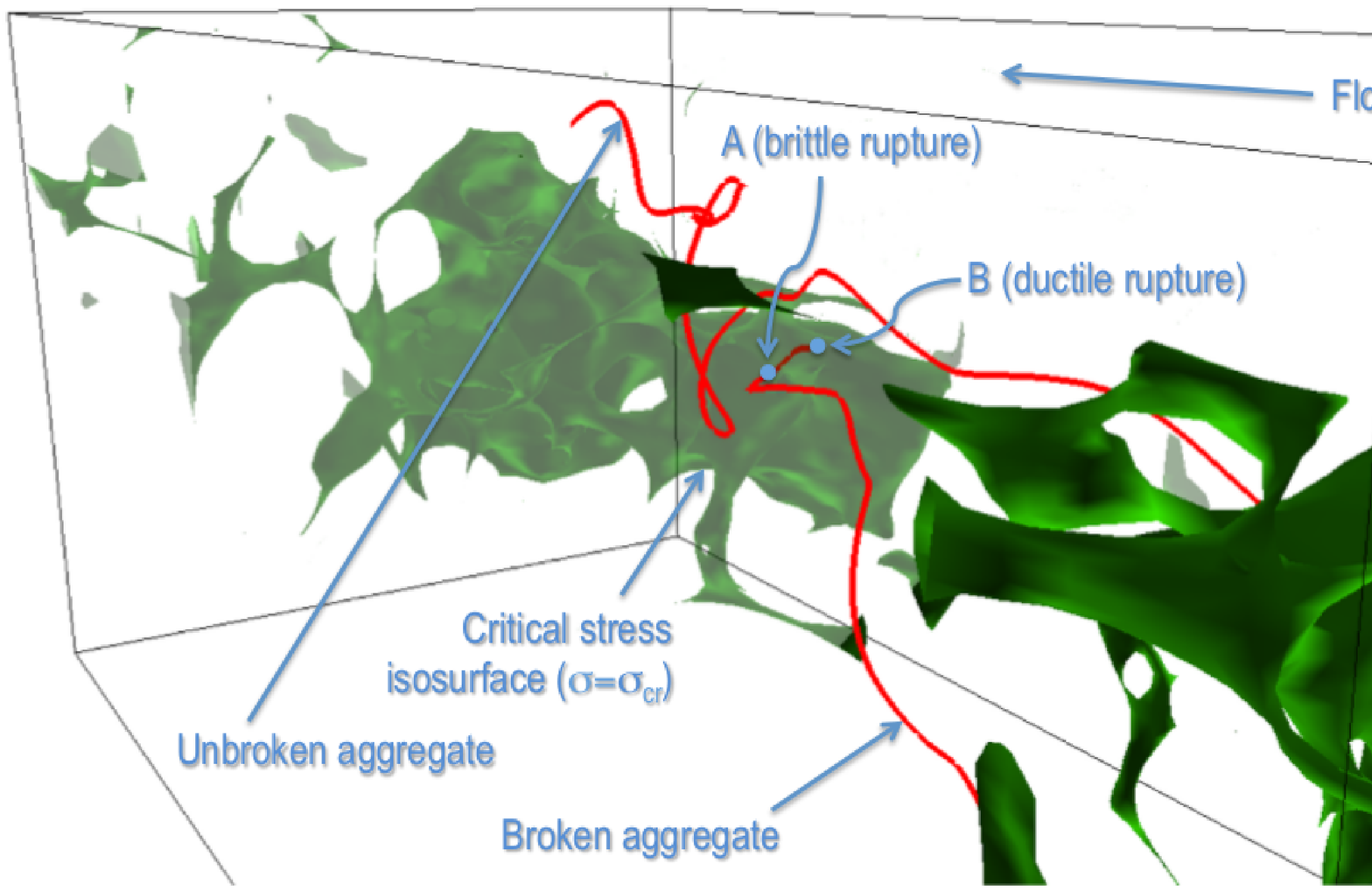}
\end{center}
\vspace{-1.0cm}
\caption{
(Color online) Rendering of brittle and ductile rupture in turbulent flow.
The trajectory of two different aggregates is shown, superimposed onto the
isosurface of the critical stress $\sigma=\sigma_{cr}$ required to produce
brittle rupture or activate ductile rupture.
The broken aggregate trespasses the $\sigma_{cr}$ isosurface at
point A (potential brittle rupture) and undergoes ductile rupture at point B (where the
breakage condition $E > E_{cr}$ is met).
The unbroken aggregate avoids all regions where $\sigma \ge \sigma_{cr}$ and does
not break within the time window considered in this figure.
Critical stress isosurface is taken at the time of ductile rupture. Aggregate trajectories
are tracked several time steps backward from this time.
}
\label{rendering}
\end{figure}

For these purposes, we assume that the
breakage process is first activated when the hydrodynamic stress acting on
the aggregate, referred to as $\sigma$ hereinafter, exceeds a critical value
that is characteristic of a given type of aggregate: $\sigma>\sigma_{cr}$
(activation condition, sufficient to produce brittle rupture).
In figure \ref{rendering}, which provides a visual rendering of the
rupture events examined in this study, this condition occurs as soon as the
aggregate trespasses the $\sigma_{cr}$ isosurface (point A along the trajectory
of the broken aggregate).
As long as the condition $\sigma>\sigma_{cr}$ is met the process continues, mimicking
the situation in which the aggregate is storing energy from the surrounding fluid. The process
comes to an end when the energy transferred from the fluid to the aggregate, referred to as
deformation energy hereinafter and defined as:
\begin{equation}
E=\int_0^{\tau} \epsilon(t|\sigma>\sigma_{cr}) \d t~~,
\label{energy}
\end{equation}
with $\tau$ being the time spent by the aggregate in regions of the flow where
$\sigma>\sigma_{cr}$ and $\epsilon$ being the dissipation rate of fluid kinetic energy,
exceeds the critical breakage value, which is also characteristic of the type of aggregate
under investigation: $E>E_{cr}$ (breakage condition).
In figure \ref{rendering}, this condition occurs at point B inside the $\sigma_{cr}$ isosurface.
In this figure, we also show the trajectory of an aggregate that avoids all regions where
$\sigma > \sigma_{cr}$ and does not break within the time window considered.
Note that the $\sigma_{cr}$ isosurface is taken at the time of ductile rupture, while aggregate
trajectories are tracked several time steps backward from this time.
To single out the effect of ductile rupture, we follow \cite{thesJFM} and assume
that the stress is $\sigma \sim \mu \left( \epsilon / \nu \right)^{1/2}$,
where $\mu$ ($\nu$) is the dynamic (kinematic) viscosity and $\epsilon = 2 \nu s_{ij} s_{ij}$,
with
$s_{ij} = \frac{1}{2} \left( \partial u_i / \partial x_j+ \partial u_j / \partial x_i \right)$
the strain rate tensor. Based on these definitions, strong fluctuations of $\epsilon$ control
the fluctuations of the stress and therefore the occurrence of breakage events.
In the limit of instantaneous breakage this translates into a picture where
an aggregate, once released into the flow, moves through it until the local dissipation exceeds
a threshold value $\epsilon_{cr}$ that causes its rupture. In this case, the main variable
to monitor is the time spent by the aggregate in regions of the flow where $\epsilon < \epsilon_{cr}$
(the so-called exit-time, which will be defined formally
in the next section).

We remark here that the proportionality between $\sigma$ and $\epsilon$ is strictly valid only for
a normal distribution for the local shear rate \cite{deli76,Pope}, a
%(Delichatsios \& Probstein, 1976; Pope, 2000), a
condition that may be violated in the near-wall region of the channel.
In addition, our approach still separates the role of turbulence from that of internal stresses, which
are neglected.
The direct coupling between hydrodynamic and internal stresses has been investigated for the first time
by \cite{lanotte-vanni} for the case of homogeneous and isotropic turbulence. Extension of the analysis
to anisotropic turbulence is currently under way and will be addressed in a subsequent paper.

%%%%%%%%%%%%%%%%%%%%%
\section{Physical Problem and Numerical Methodology}
%%%%%%%%%%%%%%%%%%%%%

The physical problem considered in this study is the dispersion of tracer aggregates
in turbulent channel flow,
which is the archetypal flow previously analyzed by this group within the
benchmark study of \cite{thesJFM}.
The flow is non-reactive, isothermal and incompressible,
and the numerical methodology is based on an Eulerian-Lagrangian approach
that has been used successfully in past investigations of turbulent dispersed flows
\cite{ms02,test-case}.
The reference geometry consists of two infinite flat parallel plates separated by
a distance $2h$. The origin of the coordinate system is located at the center
of the channel with the $x$, $y$ and $z$ axes pointing in the streamwise, spanwise,
and wall-normal directions, respectively. Periodic boundary conditions are
imposed on the fluid velocity field in the homogeneous directions ($x$ and $y$),
no-slip boundary conditions are imposed at the walls. The size of the
computational domain is $L_x \times L_y \times L_z = 4 \pi h \times 2 \pi h \times 2h$.
The shear Reynolds number is $Re_* = u_* h / \nu = 150$
(Marchioli et al., 2008), where $u_* = \sqrt{\tau_w / \rho}$ is the shear
velocity based on the mean wall shear stress. This value of $Re_*$ was chosen
to match that used by \cite{thesJFM}. We remark that, based on the findings
of \cite{sf13} and those of \cite{ham} regarding the statistical distribution of
the energy dissipation rate in turbulent channel flow up to $Re_*=600$, present
results are expected to scale up to Reynolds numbers significantly higher
than $Re_* = 150$.
All variables discussed in this paper are expressed in wall units,
obtained using $u_{*}$ and $\nu$. 

The flow solver is based on
a Fourier-Galerkin pseudo-spectral method that solves for the full Navier-Stokes
equations and thus yields
%the Fourier-Galerkin method in the streamwise and spanwise directions, and
%on a Chebishev-collocation method in the wall-normal direction. This solver
the spatial derivatives required to calculate
$\epsilon$ along the aggregate trajectory
with spectral accuracy.
Lagrangian tracking is used to calculate the trajectory of each aggregate
based on the following equation of motion: $\dot{\bf{x}}_p = {\bf{u}}_{@p}$,
with ${\bf{x}}_p$ the aggregate position and ${\bf{u}}_{@p}$ the fluid velocity
at ${\bf{x}}_p$. This equation is solved in time using a fourth-order
Runge-Kutta scheme, whereas sixth-order Lagrangian polynomials are used
to obtain the fluid velocity and the fluid velocity derivatives at the instantaneous
aggregate position.
Further details on the numerical methodology
can be found in \cite{test-case,sm09}.
Breakage measurements are performed by releasing aggregates
in two distinct regions of the channel: the wall region, which
comprises a fluid slab 10 wall unit thick where the viscous
stress (representing the mean fluid shear)
is maximum while the turbulent
stress is close to zero; and the center-plane of the bulk region,
where all wall stress contributions drop to zero and
turbulence is closer to homogeneous and isotropic.
In the following, the two release regions are labeled $\Omega_W$ and
$\Omega_C$, respectively. Within each of these regions, $2 \cdot 10^5$ aggregates were
released, their trajectories tracked and breakage events detected.

\begin{figure}
\vspace{1.0cm}
\begin{center}
\includegraphics[width=8.0cm,angle=270.]{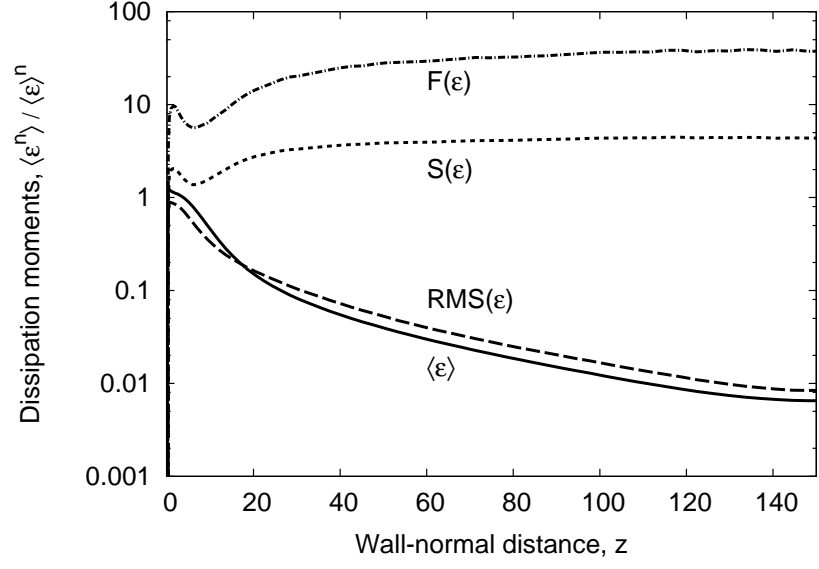}
\end{center}
\caption{
Statistical moments of the energy dissipation rate seen by the aggregates:
mean value, $\langle \epsilon \rangle$;
Root Mean Square value, $RMS(\epsilon)$;
Skewness factor, $S(\epsilon)$;
Flatness factor, $F(\epsilon)$.
Brackets indicate quantities averaged in time and space over the homogeneous
flow directions.
}
\label{edr-stats}
\end{figure}

We remark here that the dissipation of kinetic energy by viscosity in fully-developed
turbulence occurs primarily at the smallest scales of the flow, namely
at scales close of the order of the Kolmogorov scale, $\eta_K$. In wall-bounded
turbulence, the mean value of these local dissipation scales can be estimated from the relation
$\eta_K (z,t)=\left[ \nu^3 / \epsilon(z,t) \right]^{1/4}$, allowing fluctuations
in the scale to be directly connected to variations in $\epsilon$ \cite{ham}. Due to the
non-homogeneity and anisotropy of the turbulence \cite{jot2006}, dissipation results in local
values of $\epsilon$ that can be orders of magnitude larger than the mean, even for
turbulent flows at moderate Reynolds numbers like the one considered in the present study \cite{thesJFM}.
Such high amplitudes are the result of very large velocity gradients that
act on the aggregates and ultimately determine their breakage.
Variations in the smallest scales at which dissipation occurs are reflected in the statistical
moments of the energy dissipation rate, defined as
$\langle \epsilon^n \rangle / \langle \epsilon^n \rangle$, and shown
in figure \ref{edr-stats} as a function of the distance from the wall, $z$.
In this paper, angle brackets denote quantities averaged in time and in
the homogeneous directions. Note that, because we track many tracer aggregates, the
average dissipation seen by aggregates along their Lagrangian trajectory
is in practice equal to the Eulerian one,
which was investigated also by \cite{ham} at varying $Re_*$ \cite{footnote2}.
Figure \ref{edr-stats} shows that aggregates are subject to high
fluctuations of the kinetic energy dissipation even when they sample the bulk flow region.
Close to the walls, dissipation attains high mean values while fluctuations, proportional
to the Root Mean Square (RMS), are intense throughout the channel and correspond to a highly
intermittent distribution of $\epsilon$. Figure \ref{edr-stats} also shows that
the skewness ($n=3$) and flatness ($n=4$) factors,
$S(\epsilon)$ and $F(\epsilon)$ respectively, are
significantly higher than both the mean and the RMS, especially in the center
of the channel where values differ by roughly three to four orders of magnitude.
This suggests that breakage events in the bulk of the flow may be caused
by (rare) extreme energy dissipation events. However, these events are expected
to have an effect on instantaneous brittle ruptures more than on ductile ruptures,
which require the occurrence of events with a certain time persistence.
As already observed by \cite{ham}, the higher order moments agree closely with the
results in homogeneous isotropic turbulence for much of the channel, exhibiting a
universal flow-independent behavior that scales with $Re_{*}$ and is lost only in
the near-wall region.
This observation can be put in connection with the existence of scaling laws
for the breakage rate, observed for instance by \cite{thesJFM,lanotte}.
Already in the limit of brittle rupture, specific flow properties such as anisotropy and
non-homogeneity have a crucial impact on breakage dynamics since they determine
the spatial and temporal distribution of fluid stresses (and, therefore, of energy dissipation).
In wall-bounded flows such distribution exhibits features similar to homogeneous
isotropic turbulence in the bulk of the channel, where anisotropy and non-homogeneity
are not dominant: 
The behavior of the higher order moments of the energy dissipation shown in figure 1
suggests that the breakage process may exhibit universal (or nearly universal) features
only in this region. Universality is inevitably lost near the wall with important implications
for strong aggregates, which can only be broken by extreme dissipation events.

%%%%%%%%%%%%%%%%%%%%%
\section{Results and Discussion}
%%%%%%%%%%%%%%%%%%%%%

\begin{figure}

\vspace{-4.0cm}

%\begin{tabular}{cc}
\includegraphics[width=11.0cm,angle=270.]{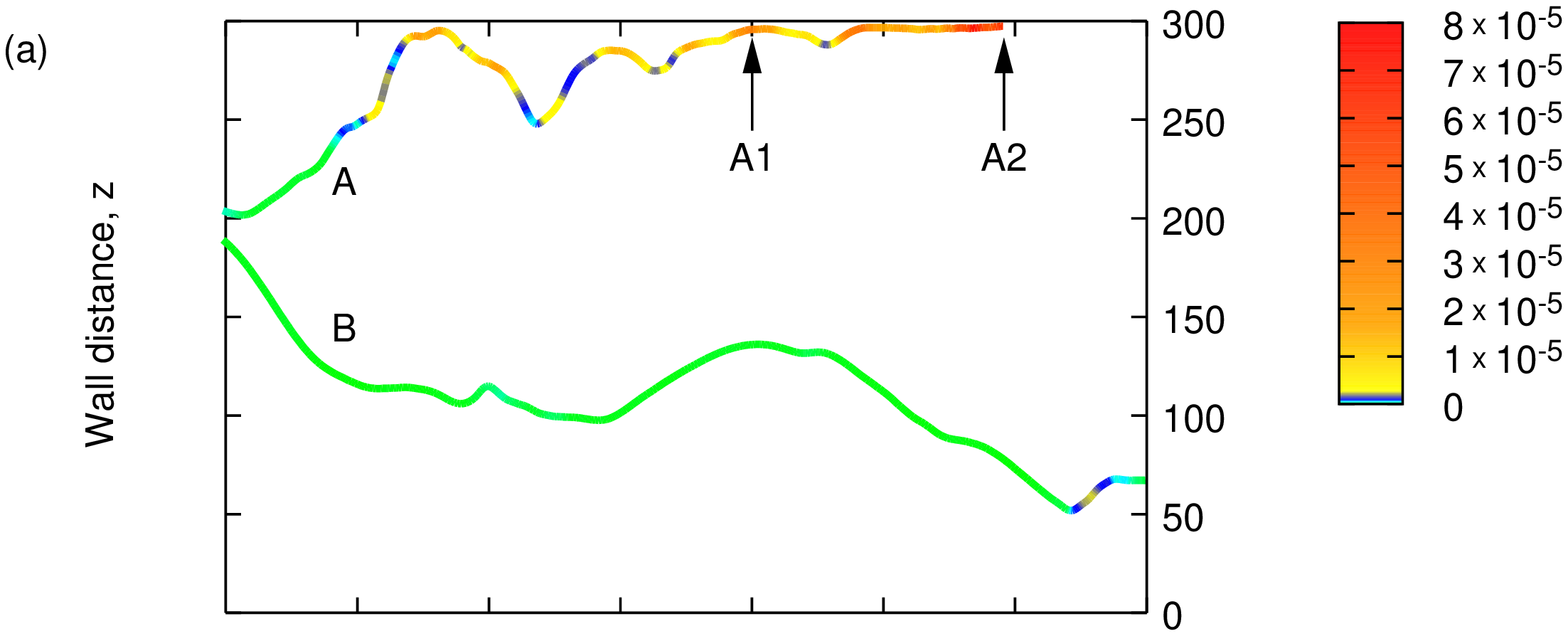}
\includegraphics[width=7.0cm,angle=270.,]{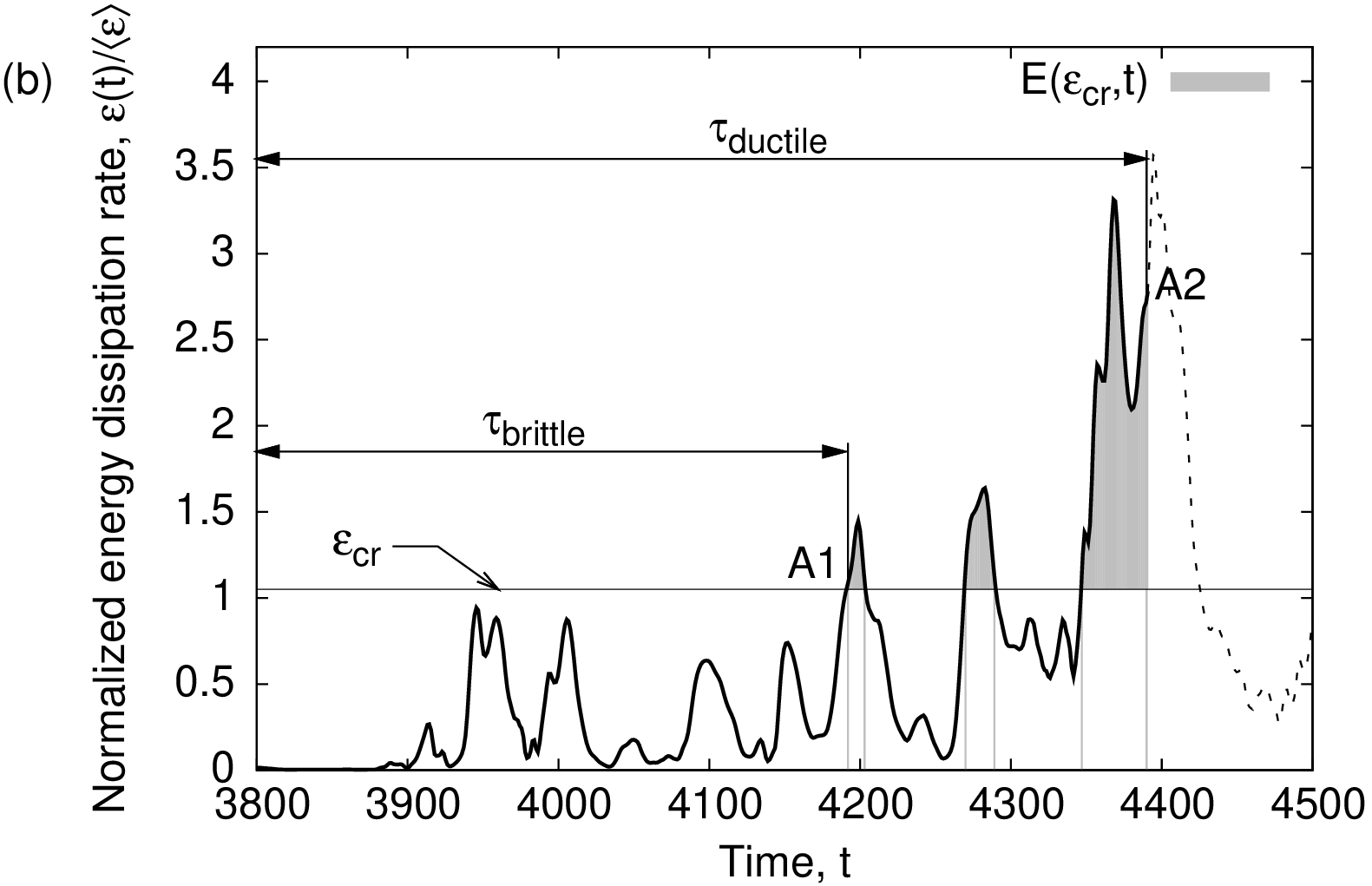}
%\end{tabular}
\vspace{-0.8cm}
\caption{
(Color online)
Time evolution of the aggregate-to-wall distance for two sample aggregates
A, subject to breakage, and B, not subject to breakage (a); and
time evolution of the energy dissipation rate seen by aggregate A along
its trajectory. The gray areas in panel (b) correspond to time windows
during which the aggregate evolves in regions of the flow where the local dissipation
is above the critical value required to either break the aggregate (in case of
brittle aggregate, $\tau_{brittle}$) or activate the breakage process (which, in case
of ductile aggregate, is brought to completion at time $\tau_{ductile}$).
The trajectories in panel (a) are colored based on the instantaneous absolute
value of the local energy dissipation rate.
Time $t$ is expressed in wall units and represents the time spent by the aggregates
within the flow after release at time $t_0$.
}
\label{time-evo}
\end{figure}

The statistics of the rupture of ductile aggregates are examined by
focusing mainly on one observable: the breakage rate.
Figure \ref{time-evo} shows schematically the procedure we followed
to estimate the breakage rate, using the trajectory of two sample
aggregates labeled as A and B.
In figure \ref{time-evo}(a), we show the time evolution of the aggregate-to-wall distance
for A and B, colored using the value (expressed in wall units) of the kinetic energy dissipation along the aggregate
trajectory: high dissipation is in dark gray (red online), low dissipation is in light gray
(blue online).
Note that $t$ represents the dimensionless time (in wall units) spent by the
aggregates within the flow after injection at time $t_0=0$. Since we are interested in the
effect of flow anisotropy on breakage, the fluid characteristic scale adopted in this study
is $\tau_f = \nu / u_*^2$.
Aggregate A is trapped at the wall and eventually breaks, aggregate B
is able to stay in the bulk of the flow during the time window of the simulation
and escapes breakage.
The procedure employed to define the breakage rate follows \cite{thesJFM,lanotte-vanni}
%Babler et al. (2012, 2015)
and can be explained with reference to the time history of the dissipation
along the trajectory of aggregate A, shown in figure \ref{time-evo}(b).
The aggregate is released at a time $t_0$ and moves within the flow for a time $t=\tau$
(referred to as exit time hereinafter) after which the local dissipation exceeds for
the first time the critical threshold $\epsilon_{cr}$, indicated by point $A1$ in figure \ref{time-evo}(b).

If the aggregate is brittle, then it breaks at point $A1$: The first crossing of
$\epsilon_{cr}$ thus defines the exit time used to compute the breakage rate: $\tau=\tau_{brittle}$
in figure \ref{time-evo}(b). The breakage rate for the given threshold is then computed as
the inverse of the mean exit time, obtained upon ensemble-averaging over many time histories
(this averaging being represented by the overbar):
\begin{equation}
f(\epsilon_{cr})=\frac{1}{ {\xoverline[1.2]{\tau}}_{brittle}(\epsilon_{cr}) }~.
\label{brittle-rate}
\end{equation}
Note that, following \cite{thesJFM,lanotte-vanni},
%Babler et al. (2012, 2015),
aggregates are released only at points where the local dissipation
is below $\epsilon_{cr}$ to avoid the occurrence of exit times
equal to zero.

\begin{figure}
\vspace{1.0cm}
\begin{center}
\includegraphics[width=8.0cm,angle=270.]{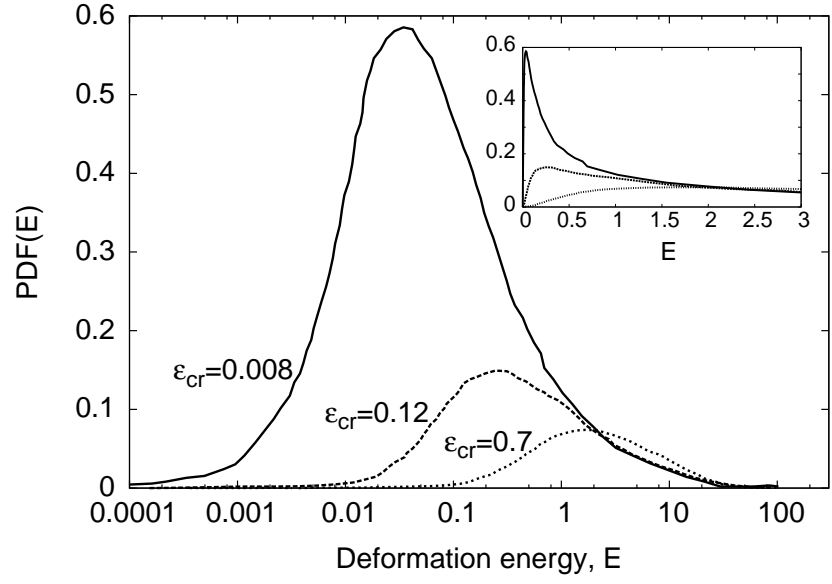}
\end{center}
\caption{
Probability density function of the deformation energy in turbulent channel flow, conditioned to the
threshold value of $\epsilon_{cr}$: $\epsilon_{cr}=0.008$ (weak aggregate, solid line);
$\epsilon_{cr}=0.12$ (mild aggregate, dashed line);
$\epsilon_{cr}=0.7$ (strong aggregate, dotted line).
}
\label{pdf-energy}
\end{figure}

If the aggregate is ductile then it does not break at $A1$, since $\epsilon > \epsilon_{cr}$
is just the activation condition. The exit time is now computed as the time required to satisfy
both the activation condition and the breakage condition, $E > E_{cr}$: in figure \ref{time-evo}(b),
the time windows during which both conditions are met encloses all the gray areas, and
breakage occurs at point $A2$ yielding an exit time $\tau=\tau_{ductile}$.
The breakage rate for the given thresholds of $\epsilon$ and $E$ is then computed as:
\begin{equation}
f(\epsilon_{cr}, E_{cr})=\frac{1}{ {\xoverline[1.2]{\tau}}_{ductile}(\epsilon_{cr}, E_{cr}) }~.
\label{ductile-rate}
\end{equation}
Note that, by definition, $E_{cr}$ depends on $\epsilon_{cr}$. This implies that the
statistical distribution of the deformation energy made available by the fluid for breakage depends
on the specific value of $\epsilon_{cr}$ that characterizes the aggregate. Such distribution is shown
in figure \ref{pdf-energy}, where we plot the Probability Density Function (PDF) of $E$ for
aggregates released in the channel center, computed according to eq. (\ref{energy})
and considering only aggregates that eventually break.
This implies that these PDFs provide also evidence of the statistical distribution
of breakage events.
The curves refer to three different values of $\epsilon_{cr}$:
$\epsilon_{cr}=0.008$, corresponding to the case of a weak aggregate for the present
flow configuration;
%with small response time to stress-induced deformation;
$\epsilon_{cr}=0.7$, corresponding to the case of a strong aggregate;
%with large response time to stress-induced deformation
and an intermediate case with $\epsilon_{cr}=0.12$, which can be referred to as mild aggregate.
As the strength of the aggregate, namely its resistance to breakage, increases, the PDF shifts toward
higher values of $E$ but exhibits lower peak values (note that PDFs are normalized such that the
area below each curve is equal to unity). This trend provides a first
characterization of the breakage events that are typically experienced by the aggregates:
very intense but relatively short in time for strong aggregates, less intense but
more persistent in time for weak aggregates.
Note that, since the horizontal axis is plotted in logarithmic scale, the PDFs deviate
from a Gaussian distribution (as shown clearly in the inset of figure \ref{pdf-energy},
where axes are in linear scale):
%Following Babler et al. (2015), 
This distribution would be obtained if breakage events were controlled mainly by
the early rupture events that occur in the vicinity of the aggregate release location
(the channel centerline) \cite{thesJFM}.
Apparently, this is not the case for ductile rupture in anisotropic turbulence.

\subsection{Breakage rates}
\begin{figure}
\vspace{1.0cm}
%\begin{tabular}{cc}
\includegraphics[width=8.0cm,angle=270.]{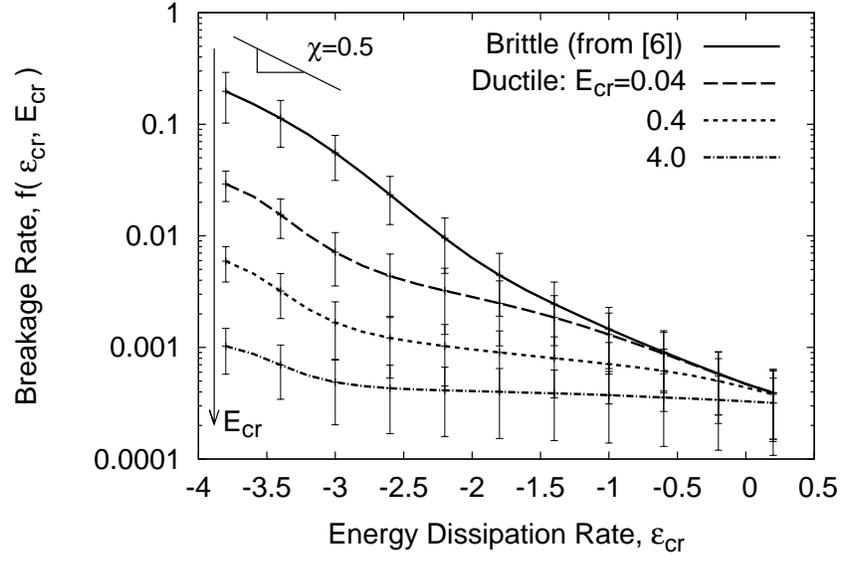}
%\end{tabular}
\caption{
Break-up rate for ductile aggregates released in the center of the channel,
$f(\epsilon_{cr},E_{cr})$. For comparison purposes, the break-up rate of brittle
aggregates (solid line, taken from \cite{thesJFM})
is shown. Error bars represent the standard deviation
from the mean value, and were computed using the variance of the exit
time, $\sigma^2_{\tau} = \langle \tau^2 \rangle - \langle \tau \rangle^2$.
}
\label{breakup-bulk}
\end{figure}

\begin{figure}
\vspace{1.0cm}
%\begin{tabular}{cc}
\includegraphics[width=8.0cm,angle=270.]{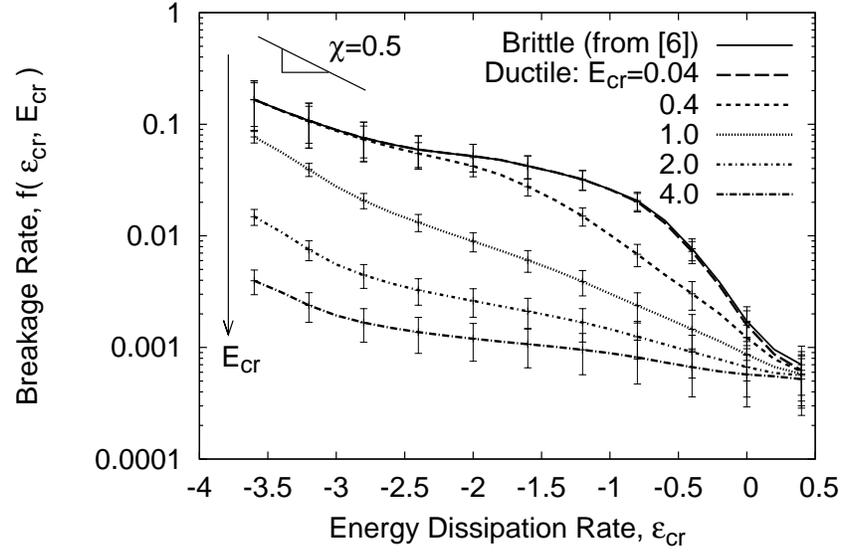}
%\end{tabular}
\caption{
Break-up rate for ductile aggregates released in the near-wall region of the
channel, $f(\epsilon_{cr},E_{cr})$. For comparison purposes, the break-up rate
of brittle aggregates (solid line, taken from \cite{thesJFM}) is shown.
Error bars are as in figure \ref{breakup-bulk}.
}
\label{breakup-wall}
\end{figure}

To examine further the breakage process, in figures \ref{breakup-bulk} and \ref{breakup-wall}
we show the rates of ductile breakage, obtained according to eq. (\ref{ductile-rate}) for the two release
locations considered in this study: $\Omega_C$ and $\Omega_W$ respectively.
These two figures show the effect of increasing the critical deformation
energy $E_{cr}$ for different threshold values of the critical energy dissipation
$\epsilon_{cr}$ and thus extend the findings of \cite{thesJFM}, which focus
on the effect of increasing $\epsilon_{cr}$ (namely the strength of a brittle
aggregate) in different flow configurations.
Results are shown for
three different values of the threshold $E_{cr}$, sampled from the distributions of
figure \ref{pdf-energy} and chosen to span two orders of magnitude:
$E_{cr}=0.04$, representing a case of low ductility for the present flow configuration;
$E_{cr}=0.4$, representing a case of intermediate ductility;
$E_{cr}=4.0$, representing a case of high ductility.
In both figures, breakage rates for brittle aggregates, computed according to eq. (\ref{brittle-rate}),
are also shown for comparison purposes.
To ensure convergence of the statistics, the breakage rates reported in figures \ref{breakup-bulk} and
\ref{breakup-wall} correspond to a percentage of broken aggregates equal to at least $80 \%$ (in the
worst case scenario of highly ductile rupture, the percentage being above $95 \%$ in all other cases).
Error bars attached to each profile represent the
standard deviation from the mean value of the breakage rate, computed
using the variance of the exit time, $\sigma_{\tau}^2 = \langle \tau^2 \rangle - \langle \tau \rangle^2$.
Error bars are shown to provide an indication of the dispersion of breakage rates
around the mean value: The higher the dispersion, the lesser the accuracy and predictive
capability of correlations based solely on $f(\epsilon_{cr},E_{cr})$.

Let us focus first on the results for aggregates released in the
center of the channel (figure \ref{breakup-bulk}).
The breakage rate of
brittle aggregates (solid curve, taken from \cite{thesJFM})
%As already shown in \cite{thesJFM}, the breakage rate of brittle aggregates (solid curve)
generally decreases with increasing
aggregate strength, in agreement with the intuitive idea that
weak aggregates in wall-bounded flows are broken by turbulent
fluctuations faster than strong aggregates.
For small $\epsilon_{cr}$ the breakage rate is known to exhibit a power-law behavior of the type
$f(\epsilon_{cr}) \propto \epsilon_{cr}^{-\chi}$,
where $\chi$ is a flow-dependent scaling exponent:
\cite{thesJFM} have demonstrated that the value of $\chi$ for aggregates released in the
central region of a channel is very similar to that of aggregates released outside a
developing boundary layer but slightly larger than that of aggregates released near the channel
walls or in homogeneous flows. In the case of figure
\ref{breakup-bulk}, the power-law scaling of $f(\epsilon_{cr})$ for brittle
aggregates is observed when $\epsilon_{cr} < -3$ and the best fit
is obtained for $\chi \simeq 0.5$.
When ductile aggregates are taken into account (dashed curves), breakage rates change
dramatically, especially for weak aggregates with low $\epsilon_{cr}$
threshold.
The values of $f(\epsilon_{cr},E_{cr})$ decrease
significantly with respect to the case of instantaneous breakage,
already at low thresholds for the critical deformation energy, $E_{cr}$
(e.g. $E_{cr}=0.04$).
In addition, no clear power-law scaling is observed anymore
and the breakage rate profiles tend to flatten as the aggregate ``ductility''
increases. As could be expected, the effect of ductile rupture on $f(\epsilon_{cr},E_{cr})$
becomes less important for strong aggregates: These must be subject to extremely
violent fluid stresses, typical of the intermittent nature of small-scale turbulence,
to activate the breakage process and thus can store
the level of energy required to break almost impulsively. As a result, there is just
a little increase of the exit time with respect to strong brittle aggregates.

It is clear from the results of figure \ref{breakup-bulk} that
ductile rupture (a process that is of course linked to restructuring phenomena
within the aggregate) has a dramatic effect on the frequency
with which small aggregates break in wall-bounded turbulence.
Any predictive model failing to reproduce this feature
would inevitably yield strong over-prediction
of the breakage rates. One example is the exponential model of
Kusters \cite{kust}, which is valid for instantaneous breakage
only and is based on the dimensional assumption
that breakage is ruled by Gaussian kinetic energy dissipation:
\begin{equation}
f(\epsilon_{cr}) = \sqrt{\frac{4\pi\epsilon}{15 \nu}} \exp{\left( - \frac{15}{2} \frac{\epsilon_{cr}}{\langle \epsilon \rangle} \right)}~.
\end{equation}
This classical model predicts a very sharp drop-off at intermediate threshold values
of $\epsilon_{cr}$ and a constant breakage rate for small threshold values, in strong
disagreement to the breakage rate found in our simulations. The discrepancy originates
both from the simplified assumption of a Gaussian dissipation and from the neglect of
ductile rupture.

The results of figure \ref{breakup-bulk} depend quantitatively on the specific location
chosen to release the aggregates at time $t_0$. In the center of the channel (release region
$\Omega_C$)
strong aggregates, no matter if subject to brittle or ductile
rupture, are mainly broken by the rare extreme
excursions of dissipation from the mean, which are caused by intermittency.
Most of such aggregates must therefore reach the high-dissipation, high-shear
regions of the flow near the channel walls to undergo breakage. To examine the influence
of the release location on breakage rates, in figure \ref{breakup-wall} we show
the behavior of $f(\epsilon_{cr},E_{cr})$ for aggregates released in the near-wall region $\Omega_W$.
Focusing first on the brittle aggregates (solid curve,
taken from \cite{thesJFM}), we observe that the power-law scaling
at small values of $\epsilon_{cr}$ is followed by a flattening for intermediate
values of the threshold, which was not observed in figure \ref{breakup-bulk}.
For the very large threshold values associated to the right end of the profile, a drop-off
in the breakage rate is observed, representing the case of aggregates that are too
strong to be broken by the mean shear alone: intense but rare turbulent fluctuations
within the near wall region are required to overcome the cohesive force of these aggregates \cite{thesJFM}.
%(Babler et al., 2015).
The inclusion of ductile rupture effects (dashed curves) produces again a clear decrease of
the breakage rates, which vanishes for large values of $\epsilon_{cr}$.
Compared to the results of figure \ref{breakup-bulk}, we observe that the decrease
is now almost negligible for aggregates with low ductility (corresponding to the $E_{cr}=0.04$
curve in figure \ref{breakup-wall}) and flattening of the profiles is only attained for
very high threshold values of the deformation energy.
We also note that error bars are generally smaller, indicating a lower variability of
the statistics: this is due to the fact that aggregates are already placed in the
high-shear regions of the flow where they preferentially break and hence sample
a reduced portion of the domain compared to aggregates released in the bulk of the
flow. In spite of these quantitative
differences, however, the reduction of $f(\epsilon_{cr},E_{cr})$ associated with ductile
rupture is evident independently of the initial position chosen to inject the aggregates
into the flow.

\subsection{Evolution of the number of aggregates}

In the previous section, we pointed out that strong aggregates can move away from the
location of release and travel towards the high shear regions close to the walls.
Obviously, this dynamics has an influence on the breakage process, which is also reflected
in the time evolution of the number of unbroken aggregates, $N_{\epsilon_{cr}}(t)$.
This quantity can be used to derive the following approximation for the breakage rate of brittle aggregates that
is valid when breakage is driven by homogeneous and temporally uncorrelated stresses \cite{lanotte-vanni}:
%(Babler et al., 2012):
%
\begin{equation}
f(\epsilon_{cr})= - \frac{\d \ln N_{\epsilon_{cr}}(t)}{\d t}~,
\label{proxy}
\end{equation}
where $N_{\epsilon_{cr}}(t)$ can be simply linked to the exit time
by the relation:
\begin{equation}
\frac{N_{\epsilon_{cr}}(t)}{N(t_0)} = 1 -
\int_0^{\tau} p_{\epsilon_{cr}}(\tau) \d \tau~,
\label{proxy2}
\end{equation}
with $N(t_0)$ the number of aggregates initially released into
the flow and $p_{\epsilon_{cr}}(\tau)$ the PDF of the exit time
for a given threshold $\epsilon_{cr}$.
Based on eq. (\ref{proxy2}), $N_{\epsilon_{cr}}(t)$ is proportional
to the cumulative exit time distribution \cite{lanotte-vanni}.
%(Babler et al., 2012).

\begin{figure}
\vspace{-1.0cm}
\hspace{-2.0cm}
%\begin{tabular}{cc}
%%\includegraphics[width=8.0cm,angle=270.]{Evolution_unbroken_aggregates-Z150-BULK-linear_scale.ps}
\includegraphics[width=19.0cm,angle=0.]{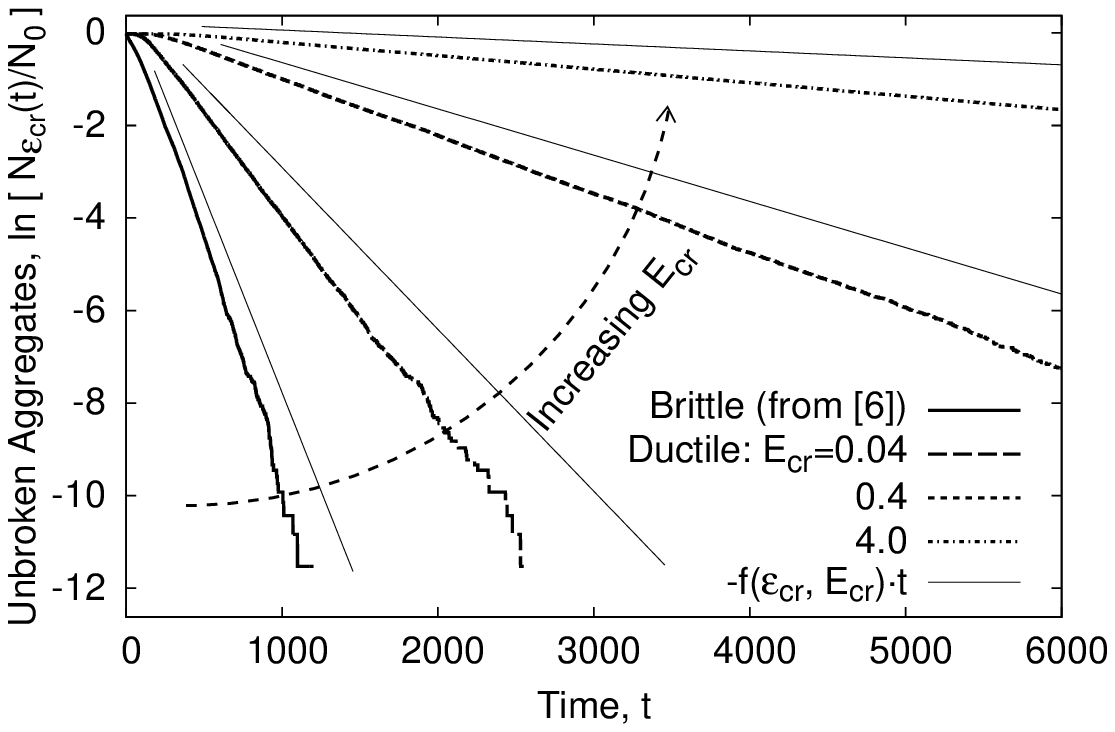}
%\end{tabular}
\vspace{-17.5cm}
\caption{
Time evolution of the number of unbroken aggregates (for brittle and ductile rupture of aggregates
released in the center of the channel). For each type of rupture, the exponential decay predicted by
 eq. (\ref{proxy}) when $N_{\epsilon_{cr}}(t) \simeq N(t_0) \exp [ -f(\epsilon_{cr}) \cdot t ]$ (thin
 solid segments) is shown. Solid line taken from \cite{thesJFM}.
}
%%\vspace{-8.95cm}
%%\hspace{-0.44cm}
%%\begin{tikzpicture}
%%   \draw [black,thick,dashed] arc (270:349:5cm);
%%\end{tikzpicture}
%%
\label{unbroken-bulk}
\end{figure}

In figure \ref{unbroken-bulk}, the evolution of the number of
aggregates released in the centerline of the channel is reported.
In particular the figure shows the behavior of
$\ln \left[ N_{\epsilon_{cr}}(t)/N(t_0) \right]$, considered here because
the corresponding slope provides a direct estimate of the aggregate
breakage rate, as suggested by eq. (\ref{proxy}).
The different curves refer
to a reference threshold $\epsilon_{cr}=0.008$ and to different
types of aggregate rupture: brittle (solid line taken from \cite{thesJFM}),
weakly ductile (dashed line), mildly ductile (dotted line) and
highly ductile (dash-dotted line).
The lowest threshold value for $\epsilon_{cr}$ was chosen because we know from \cite{thesJFM}
%Babler et al. (2015)
that in this limit the number of unbroken aggregates
decays exponentially as
$N_{\epsilon_{cr}}(t) \simeq N(t_0) \exp [ -f(\epsilon_{cr}) \cdot t ]$,
yielding $\ln \left[ N_{\epsilon_{cr}}(t)/N(t_0) \right] \simeq -f(\epsilon_{cr}) \cdot t$
with deviations due only to statistical noise (e.g. at late times when
the number of aggregates has become very small). The behavior of
$-f(\epsilon_{cr}) \cdot t$ for the three types of ductile rupture
examined in figure \ref{unbroken-bulk} is represented by the thin solid
lines.

The results of figure \ref{unbroken-bulk} show that, in general,
the evolution of the number of aggregates follows an exponential
decay only at short and intermediate times, in good agreement with
the linear segments of $-f(\epsilon_{cr}) \cdot t$. Then the decay
turns into a faster decrease at later times, associated to clear
deviations from $-f(\epsilon_{cr}) \cdot t$.
These deviations are particularly evident for ductile rupture,
indicating that simple estimates like the one given in eq. (\ref{proxy})
do not provide a reasonable approximation of the breakage rate
anymore, in agreement with the findings of figure 3.

%%%%%%%%%%%%%%%%%%%%%
\section{Conclusions}
%%%%%%%%%%%%%%%%%%%%%

In this work, we examine the breakage of small colloidal aggregates in
non-homogeneous anisotropic turbulence. In particular we focus on the
breakage rate of massless aggregates that are subject to ductile rupture caused
by the hydrodynamic fluid stresses acting on the aggregate.
This process is activated when the fluctuating hydrodynamic stress
generated by the surrounding fluid exceeds a critical value characteristic
of a given type of aggregate,
$\sigma>\sigma_{cr}$, and ends when the energy given up to the aggregate
by the surrounding fluid exceeds the critical breakage value.
From a physical point of view, the process of ductile breakage comes closer
to real applications compared to the case of instantaneous rupture considered
in previous works \cite{thesJFM}.
To compute
the breakage rate statistics, breakage kinetics under a realistic set
of assumptions are explored by means of direct numerical simulation of
turbulent channel flow seeded with a large number of aggregates,
modelled as sub-Kolmogorov massless point-particles.
Results show that the effects associated to ductile rupture
are important and lead to strong reductions of the breakage rate
with respect to instantaneous rupture.
The mechanism of ductile breakage thus acts
as a low-pass filter for stress-induced events that occur at
time scales shorter than the characteristic time with which the aggregate
responds to deformation.
The reduction in the breakage rates is evident especially for weak aggregates
characterized by small critical stress value and no universal scaling can be observed.
For strong aggregates characterized by large critical stress value the breakage rate is
less affected by the specific mechanism leading to rupture because such
aggregates can only be disrupted by extremely intense stresses and thus store
the amount of deformation energy required to break almost impulsively.

Future investigations will try to evaluate if there is
a preferred direction along which breakage takes place.
In the recent experimental measurements of aggregate breakage in laminar
and turbulent shear flows, performed by \cite{saha}, it was found that
aggregates tend to break in the direction along which they experience
the maximum stretching. Our aim is to verify the
persistence of this behavior in presence of anisotropic turbulence.

%%%%%%%%%%%%%%%%%%%%%%%%%%
\section*{Acknowledgments}
%%%%%%%%%%%%%%%%%%%%%%%%%%

COST Action FP1005 ``Fiber suspension flow modelling: a key for innovation and competitiveness
in the pulp and paper industry'' is gratefully acknowledged. Financial support from Regione
Friuli Venezia Giulia under the research project ``Sviluppo di filtri catalitici e antiparticolato ad alta
efficienza per una sostenibile mobilit\a compatibile con Euro 6'' is also highly appreciated.
A.S. is thankful to Prof. H. Pitsch for his insightful suggestions. The authors thank
Dr. Marco Svettini for running some of the simulations.

\clearpage
\newpage

%%%%%%%%%%%%%%%%%%%%%%%%%%%%%%%%%%%%%%%%%%%%%%%%%%%%%%%%%%%%%%%%%%%%%%%%%%%%%%%%%%%%%%%%%%%%%%
%                                           FIGURES                                          %
%%%%%%%%%%%%%%%%%%%%%%%%%%%%%%%%%%%%%%%%%%%%%%%%%%%%%%%%%%%%%%%%%%%%%%%%%%%%%%%%%%%%%%%%%%%%%%

\clearpage
\newpage

\clearpage
\newpage

\clearpage
\newpage

\clearpage
\newpage

\clearpage
\newpage

\clearpage
\newpage

\clearpage
\newpage

\end{document}